\documentclass{sf2a-conf2018}
\usepackage{graphicx}
\usepackage{hyperref}
\usepackage[]{natbib}  
\usepackage{epstopdf}
\usepackage{array}
\usepackage{slashbox}
\usepackage{colortbl}
\usepackage{xcolor}
\usepackage{multicol}

\def\BibTeX{{\rm B\kern-.05em{\sc i\kern-.025em b}\kern-.08em
    T\kern-.1667em\lower.7ex\hbox{E}\kern-.125emX}}
\bibpunct{(}{)}{;}{a}{}{,}  

\begin{document}

\TitreGlobal{SF2A 2017}


\title{FliPer: Classifying TESS pulsating stars}

\author{L. Bugnet$^{1,}$}\address{IRFU, CEA, Universit\'e Paris-Saclay, F-91191 Gif-sur-Yvette, France}\address{AIM, CEA, CNRS, Université Paris-Saclay, Université Paris Diderot, Sorbonne Paris Cité, F-91191 Gif-sur-Yvette, France}
\author{R. A. Garc{\'{\i}}a$^{1,2}$}
\author{G. R. Davies$^{3,}$}\address{School of Physics and Astronomy, University of Birmingham, Edgbaston, Birmingham, B15 2TT, UK}\address{Stellar Astrophysics Centre, Department of Physics and Astronomy, Aarhus University, Ny Munkegade 120, DK-8000 Aarhus C, Denmark}
\author{S. Mathur$^{5,6,}$}\address{Universidad de La Laguna, Dpto. de Astrof\'{\i}sica, E-38205, La Laguna, Tenerife, Spain}\address{Instituto de Astrofísica de Canarias, E-38200, La Laguna, Tenerife, Spain}\address{Space Science Institute, 4750 Walnut Street Suite 205, Boulder, CO 80301, USA}
\author{O. J. Hall$^{3}$}
\author{B. M. Rendle$^{3}$}

\date{Received / Accepted}

\setcounter{page}{237}

\maketitle

\begin{abstract}
The recently launched NASA Transiting Exoplanet Survey Satellite (TESS) mission is going to collect lightcurves for a few hundred million of stars and we expect to increase the number of pulsating stars to analyze compared to the few thousand stars observed by the CoRoT, \textit{Kepler} and K2 missions. However, most of the TESS targets have not yet been properly classified and characterized. In order to improve the analysis of the TESS data, it is crucial to determine the type of stellar pulsations in a timely manner. We propose an automatic method to classify stars attending to their pulsation properties, in particular, to identify solar-like pulsators among all TESS targets. It relies on the use of the global amount of power contained in the power spectrum (already known as the FliPer method) as a key parameter, along with the effective temperature, to feed into a machine learning classifier. Our study, based on TESS simulated datasets, shows that we are able to classify pulsators with a $98\%$ accuracy.
\end{abstract}

\begin{keywords}
asteroseismology - methods: data analysis - stars: oscillations
\end{keywords}


\section{Introduction}
The NASA Transiting Exoplanet Survey Satellite (TESS) conducts a nearly all-sky photometric survey providing observations for more than $400$ million stars with a $30$ minutes observational cadence from the analysis of the full frame images \citep{2014SPIE.9143E..20R}. Even if the main purpose of the TESS mission concerns small planets ($R < 4R_T$) detection and future characterization with complementary ground-based observations, the observational conditions are good enough to perform asteroseismology on bright ($T_{mag} < 15$) Solar-like pulsating TESS targets \citep{2016arXiv161006460C}. Asteroseismology has already shown high performance when providing precise estimates of mass and radius for $\sim 20,000$ Solar-type pulsators observed by the CoRoT, \textit{Kepler} and K2 missions. Asteroseismology has also proved its ability to infer precise ages for stars in the Milky way  \citep[e.g.][]{2014arXiv1409.2267M}. However, some \textit{Kepler} and most K2 targets still wait to be classified among Solar-type pulsators, classical pulsators, etc. For instance \cite{2016ApJ...833..294M} showed that more than $\sim 1000$ new red giants have been discovered as misclassified among \textit{Kepler} data, 5 years after the end of the \textit{Kepler} main mission. It demonstrates that it usually takes a large amount of time to classify stars, and it is a requirement to ensure the completeness of any set of stars to be used in any galactic population studies. We are thus looking for automatic classification methods for future missions such as the NASA TESS mission in order to provide a more accurate classification of targets to the community in a shorter time lapse. In this first attempt, we mostly focus on distinguishing solar-like pulsators (from the main sequence to the red-giant branch) from classical pulsators.\\

\cite{2011ApJ...741..119M} pointed out the dependency of granulation with the age of the star for Solar-type pulsators. This dependency can be used in the time domain by directly measuring Flicker \citep{2016ApJ...818...43B, 2013Natur.500..427B} but \citep{2018arXiv180905105B} showed that the full potential of such a dependency can be better exploited in the Fourier domain. Indeed, a star can be characterized by the amount of power in different frequency ranges in the power spectrum. While a solar-type pulsator shows relatively low oscillation amplitude components, RRLyraes and Cepheids present high peaks of power at evenly spaced frequencies that completely dominates their power spectrum.\\

\begin{figure}[ht!]
 \centering
 \includegraphics[height=12cm,clip,angle=270]{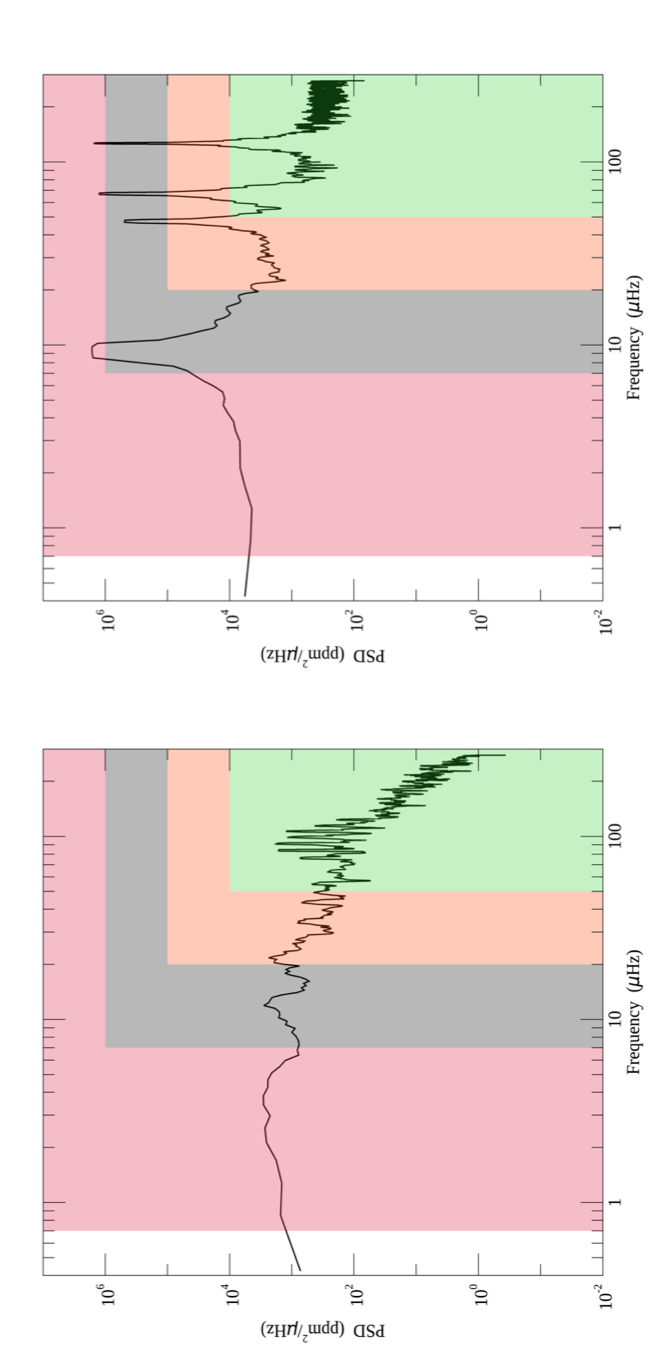}  
  \caption{\textbf{Left:} Simulated power spectrum density of a Solar-like star observed by TESS ($T_{eff}=4743 K$, $\log g=2.83 dex$). \textbf{Right:} Same for a simulated $\delta$-Scuti/$\gamma$-Dor hybrid pulsator ($T_{eff}=8578 K$, $\log g=3.89 dex$). Coloured areas (resp. red, grey, orange and green) represent the different ranges of frequency used for FliPer calculation (from resp. $0.7$, $7$, $20$ and $50$ $\mu$Hz to the Nyquist frequency).}
  \label{Fig1}
\end{figure}

FliPer was first developed as a method to automatically estimate surface gravities of Solar-type pulsators from 0.3 to 4.5 dex \citep{2018arXiv180905105B} from the global amount of power contained in their power spectrum density. In this work we modify the FliPer methodology and apply an improved version of the FliPer metric to all type of stars, including solar-like and so called classical pulsators, in order to automatically distinguish between the different spectral types. \\

\section{Methods}
\subsection{Data preparation}
 
We use $20,000$ TESS simulated light curves to produce the same number of power spectrum density representing all of the categories of star we want to classify. The classes of pulsators we consider are listed below:
\begin{multicols}{5}
\begin{itemize}
    \item Solar-like
    \item sdBV
    \item $\beta$-Cephei
    \item SPB
    \item $\delta$-Scuti
    \item $\gamma$-dor
    \item roAp
    \item RRLyrae
    \item LPV
    \item Cepheid    
\end{itemize}
\end{multicols}

Simulated dataset are taken from the work of the TESS Data for Asteroseismology (T'DA) working group \citep{2017EPJWC.16001005L} of the TESS Asteroseismic Science Operations Center (TASOC). Data can be downloaded after registration on the TASOC website\footnote{\url{https://tasoc.dk/wg0/SimData}} . We present the work made with the ``clean'' dataset, which only includes stellar signal, by opposition with the ``noisy'' data that also includes photometric noise. The data set is randomly split into a ``training set'' representing $80\%$ of the total amount of stars in the sample and a ``test set'' that contains the remaining $20\%$ of stars.

\subsection{FliPer measure}

The FliPer method \citep{2018arXiv180905105B} was first developed as a tool to estimate global parameters  of Solar-like pulsators, such as the frequency of the mode's envelope maximum power ($\nu_{max}$) or surface gravities. It relies on the measurement of global power contained in the power density spectrum of the star. Given the way to compute the FliPer it is sensitive to different variabilities present in the lightcurves: granulation, rotation, and modes for solar-like pulsators. As all these components vary when the star evolves, the FliPer value gives constraints on the evolutionary stage of the solar-like pulsator \citep{2018arXiv180905105B}. We define FliPer as: \begin{equation}
   \textnormal{F}_{\textnormal{p}} = \overline{\textnormal{PSD}} - \textnormal{P}_\textnormal{n} ,
   \label{powvar}
\end{equation}
     
where $\overline{\textnormal{PSD}}$ represents the averaged value of the power spectrum density from a giving frequency to the Nyquist frequency and $\textnormal{P}_\textnormal{n}$ is the photon noise (see \cite{2017sf2a.conf...85B} for more information). \\

\subsection{FliPer values calculation}

For each star we calculate different values of FliPer corresponding to four different frequency ranges denoted by $\textnormal{F}_{\textnormal{p},k}$ where $k =[0.7, 7, 20, 50]$ $\mu$Hz corresponding to a starting frequency of $0.7$, $7$, $20$, and $50$ $\mu$Hz as represented by the coloured areas on Fig.~\ref{Fig1}. By combining these different FliPer values, we have access to different parts of the power spectrum density of the star (see \cite{2018arXiv180905105B} for more details about the method).\\

By calculating these FliPer values for all kind of stars, it is possible to classify pulsators depending on the amount of power contained in their power spectra. Indeed, while a solar-type pulsator shows relatively low oscillation amplitude components, RRLyraes and Cepheids present high peaks of power at evenly spaced frequencies that completely dominate their power spectrum. This would result in completely different FliPer values ranging on several order of magnitudes. Figure~\ref{Fig1} shows the power spectrum density for a solar-like star ({Left} panel) and a $\delta$-Scuti/$\gamma$-Doradus hybrid pulsator ({Right} panel). We can clearly observe that the classical pulsator shows much larger oscillation peaks than the solar-like star. The nature of the star affects each value of FliPer, leading to a characteristic pattern of FliPer values $\textnormal{F}_{\textnormal{p},k}$, corresponding to each type of pulsator. \\

\subsection{Classification algorithm}

Instead of determining ourselves the different FliPer patterns associated with each type of stars, we decided to use a random forest classifier \citep{Breiman2001} in order to classify the stars. The algorithm constructs decision trees during the training, and combines them to automatically get the most probable class for each star of the test set. This is done by using the ``RandomForestClassifier'' function from the ``sklearn.ensemble'' Python library \citep{scikit-learn}. The input parameters are $\textnormal{F}_{\textnormal{p} 0.7}$, $\textnormal{F}_{\textnormal{p} 7}$, $\textnormal{F}_{\textnormal{p} 20}$, $\textnormal{F}_{\textnormal{p} 50}$ and $T_{eff}$. The algorithm learns on the ``true'' classes and gives as output variable the ``predicted'' classes.

\section{Results \& Conclusion}
By applying our methodology based on a random forest algorithm, the classes are very well reconstructed: we obtain a $98\%$ accuracy on the classification of the test set with our algorithm. Considering that the FliPer method was first built to analyze physical properties of Solar-type pulsators, this is not a surprise to see misclassified classical pulsators. We also point out that more than $99\%$ of the real Solar-like pulsators are well classified by the algorithm. The next step would be to use the FliPer regressor \citep{2018arXiv180905105B} that has already been applied to \textit{Kepler} targets to estimate surface gravities of newly classified TESS solar-like pulsators. This very good result will allow us to study most Solar-like pulsators observed by TESS, once the algorithm has been re-trained with real TESS data. FliPer is being integrated in the TASOC Stellar Classification module, inside a larger Random Forest classifier that will be used to automatically classify TESS targets. The pipeline (Tkatchenko et al., \textit{in prep}) also includes many other methods of classification such as clustering algorithms, or deep learning convolution networks \citep[e.g.][]{2018MNRAS.tmp..471H}.

\begin{acknowledgements}
We thank the T'DA team for useful discussion and remind the reader that this algorithm is part of the classification pipeline used for TESS classification from TASOC. L.B. and R.A.G. acknowledge the support from PLATO and GOLF CNES grants.
S.M. acknowledges support by the National Aeronautics and Space Administration under Grant NNX15AF13G, by the National Science Foundation grant AST-1411685 and the Ramon y Cajal fellowship number RYC-2015-17697.
B.M.R. acknowledge the support of the UK Science and Technology Facilities Council (STFC). Funding for the Stellar Astrophysics Centre is provided by the Danish National Research Foundation (Grant DNRF106).
\end{acknowledgements}

\bibliographystyle{aa}  


%
\end{document}